\RequirePackage{fix-cm}
\documentclass[smallextended]{svjour3}       
\smartqed  
\usepackage{amssymb}
\usepackage{amsmath}
\usepackage{color}
\usepackage{graphicx}
\usepackage[justification=centering]{caption}
\usepackage{multirow}
\usepackage{extarrows}
\usepackage{booktabs}
\usepackage{bigstrut}
\begin{document}

\title{An efficient and secure arbitrary $N$-party quantum key agreement protocol using Bell states
}


\author{Wen-Jie Liu     \and
        Yong Xu         \and
        Ching-Nung Yang  \and
        Pei-Pei Gao     \and
        Wen-Bin Yu     
}


\institute{
           W.-J. Liu \and
           W.-B. Yu
           \at Jiangsu Collaborative Innovation Center of Atmospheric Environment and Equipment Technology (CICAEET), Nanjing 210044, P.R.China
           \at Jiangsu Engineering Center of Network Monitoring, Nanjing University of Information Science \& Technology, Nanjing 210044, P.R.China
           \\
           \email{wenjiel@163.com}
           \and
           W.-J. Liu   \and
           Y. Xu       \and
           P.-P. Gao  \and
           W.-B. Yu
           \at School of Computer and Software, Nanjing University of Information Science \& Technology, Nanjing 210044, P.R.China
           \and
           C.-N. Yang
           \at Department of Computer Science and Information Engineering, National Dong Hwa University, Hualien 974, Taiwan
 }

\date{Received: April 18, 2017 / Accepted: date}

\maketitle

\begin{abstract}
Two quantum key agreement protocols using Bell states and Bell measurement were recently proposed by Shukla et al. [\emph{Quantum Inf. Process}. 13, 2391(2014)]. However, Zhu et al. pointed out that there are some security flaws and proposed an improved version[\emph{Quantum Inf. Process}. 14, 4245(2015)]. In this study, we will show Zhu et al.'s improvement still exists some security problems, and its efficiency is not high enough. For solving these problems, we utilize four Pauli operations $\left\{I, Z, X, Y\right\}$ to encode two bits instead of the original two operations $\left\{I, X\right\}$ to encode one bit, and then propose an efficient and secure arbitrary $N$-party quantum key agreement protocol. In the protocol, the channel checking with decoy single photons is introduced to avoid the eavesdropper's flip attack, and a post-measurement mechanism is used to prevent against the collusion attack. The security analysis shows the present protocol can guarantee the correctness, security, privacy and fairness of quantum key agreement.

\end{abstract}

\section{Introduction}
\label{intro1}
In recent decades, with the theory of quantum mechanics utilized in the information processing field, quantum cryptography communication has aroused more and more researchers' attention, and many important research findings are presented, including quantum key distribution (QKD)[1, 2], quantum secure sharing(QSS)[3, 4], quantum secure direct communication(QSDC)[5-9], quantum private comparison(QPC)[10-13], quantum sealed-bid auction(QSBA)[14, 15], quantum remote state preparation(QRSP)[16,17], quantum key agreement(QKA)[18, 19], and so on.

QKA is an important research topic in cryptography, which allows participants to negotiate a classical shared secret key via public quantum channels. Furthermore, each participant in QKA equally contributes to the generation of a shared key, and the shared key cannot be completely decided by any nontrivial subset of the participants. It is obviously different from QKD, where one participant decides the shared key and distributes it to other parties. So, QKA shows much more fair and secure in the generation of multi-party's shared keys than QKD. Precisely, a secure QKA protocol should meet the four properties[20] :

1) \textbf{\emph{Correctness}} Each participant involved in the protocol could get the correct shared key;

2) \textbf{\emph{Security}} An outside eavesdropper cannot get any useful information about the final shared key without being detected;

3) \textbf{\emph{Privacy}} Each participant in the protocol cannot learn any useful information about other participant's secret keys;

4) \textbf{\emph{Fairness}} All involved participants are entirely peer entities and can equally influence the final shared key, i.e., the subset of the participants cannot succeed in determining the shared key alone.

The pioneering QKA protocol was proposed by Zhou et al.[18] in 2004, which used quantum teleportation technique to generate a secret key. However, Tsai and Hwang[21] pointed out that a party in Zhou et al.'s protocol can fully determine the shared key alone without being detected and provided an improved version. Since then, a large number of QKA protocols use different quantum resources have been proposed, such as single photons[19], Bell states[22, 23], GHZ states[24, 25], cluster states [26, 27].

Recently, Shukla et al.[28] proposed two QKA protocols(SAPs for short) based on Bell states and Bell measures, the first one is designed for two parties(SAP1), and the second one is for three parties (SAP2). In the encoding phase of the protocols, each participant applies two kinds of unitary operations $\left\{ {I,{\rm{ }}X} \right\}$ and each operation represents one classical bit. However, Zhu et al.[20] found that Shukla et al.'s protocols are not secure. In specific, the privacy and fairness cannot be guaranteed in SAP2, i.e., each user can deduce the other two participants' encoding information, and any two dishonest participants can collaborate to decide the shared key alone. In addition, there is a correctness flaw in both SAP1 and SAP2, i.e., the eavesdropper can flip any bit of the final key without being detected (namely the eavesdropper's attack). Finally, they improved SAP2 by adding some additional Pauli operations to solve the flaws. However, the improvement is not efficient enough, while there still exist the privacy and fairness problem in SAP1. In order to enhance the efficiency, this study aims to propose an efficient arbitrary $N$-party QKA protocol by applying four Pauli operations $\left\{I, Z, X, Y\right\}$ for encoding key information. In the protocol, the single photon checking strategy and a post-measurement mechanism are introduced to guarantee the security. Obviously, the $N$-party QKA protocol is a general pattern, which can be considered as the improvement of two-party SAP protocol(i.e., SAP1) when $N$=2 and three-party SAP protocol(i.e., SAP2) when $N$=3.

The rest of this paper is organized as follows, in the next section, we review the two-party and three-party SAP protocols, as well as Zhu et al.'s improvement. In Sect 3, we pointed out the problems of Zhu et al.'s improvement, and then propose an efficient and secure arbitrary $N$-party QKA protocol. Finally, efficiency and security analysis are discussed in detail in the last section.

\section{Review of SAPs and Zhu et al.'s improvement}
\label{sec:2}
\subsection{Review of SAP1 and SAP2}
\label{sec:2.1}
In SAP1, Two participants Alice and Bob hold their secret keys: ${K_A} = {\kern 1pt} \{ K_A^1,{\kern 1pt} {\kern 1pt} K_A^2,{\kern 1pt} K_A^3,...,K_A^n\} $, ${K_B} = {\kern 1pt} \{ K_B^1,{\kern 1pt} {\kern 1pt} K_B^2,{\kern 1pt} K_B^3,...,K_B^n\}$, and they want to equally generate the final key $K$. The whole procedures can be briefly described as follows (also shown in Fig. 1).

\textbf{Step 1}. Alice prepares $n$ Bell states $\left| {{\psi ^{\rm{ + }}}} \right\rangle  = \frac{{\left| {00} \right\rangle  + \left| {11} \right\rangle }}{{\sqrt 2 }}$, and divides the first and the second qubits to form two sequences ${p_A}$ and ${q_A}$, respectively.

\textbf{Step 2}. Alice additionally prepares $n$/2 Bell states $\left| {{\psi ^ + }} \right\rangle $ as decoy qubits and concatenates them with ${q_A}$ to get sequence ${q'_A}$. Subsequently, Alice applies permutation operator $({\rm\Pi}_{2n})_A$ on ${q'_A}$ to create ${q''_A}$ and then sends them to Bob.

\textbf{Step 3}. After bob sends the authentic acknowledge of the receipt to Alice, Alice announces the coordinates $({\rm\Pi}_{2n})_A$. Thus Bob rearranges the qubits and performs the Bell measurements on the decoy particles, and then computes the error rate. The protocol will continue only if the error rate is within tolerable limit, otherwise it goes back to Step 1.

\textbf{Step 4}. Bob drops the decoy qubits to obtain ${q_A}$, then he encodes ${K_B}$ by applying unitary operation on each qubit of the sequence ${q_A}$. For example, to encode $K_B^i = 0\left( 1 \right)$, he applies $I(X)$ on $q_A^i$. This forms a new sequence ${q_B}$. Subsequently, Bob concatenates ${q_B}$ with $n/2$ Bell states $\left| {{\psi ^ + }} \right\rangle $ and applies the permutation operator $({\rm\Pi}_{2n})_B$ to obtain new sequence ${q'_B}$ which he sends to Alice.

\textbf{Step 5}. After receiving ${q'_B}$, Alice checks the channel as same as the procedure in Step 3. If the error rate is within the threshold, the protocol continues the next step, otherwise it goes back to Step 1.

\textbf{Step 6}. Alice publicly announces her key ${K_A}$ and Bob combines it with ${K_B}$ to generate the shared key: $K = {K_A} \oplus {K_B}$.

\textbf{Step 7}. Bob announces the order of the message qubits $\left( {{\rm\Pi _n} \subset {\rm\Pi _{2n}}} \right)$, and Alice uses the information to obtain ${q_B}$. After combing ${p_A}$ and ${q_B}$, Alice performs the Bell measurements on $p_A^iq_B^i$. and then reveals ${K_B}$ by comparing the measurement results and the initial states.

\textbf{Step 8}. With ${K_A}$ and ${K_B}$, Alice can obtain the final secret key $K = {K_A} \oplus {K_B}$.

\begin{figure}
  \centering
  \includegraphics[width=6cm]{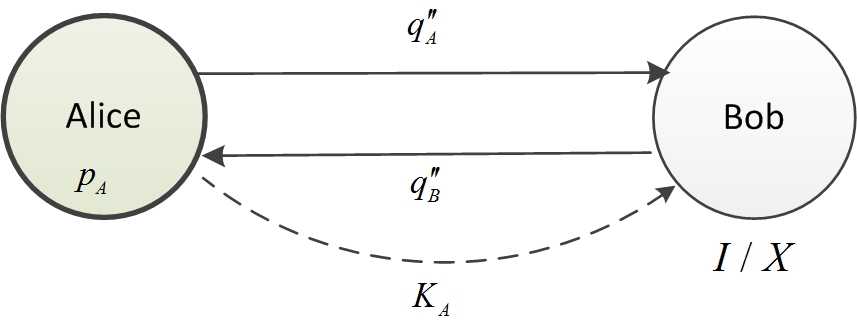}
  \caption{ The process of Shukla's two-party QKA protocol(SAP1).}
\end{figure}

 In Shukla's three-party QKA protocol (i.e., SAP2), Alice, Bob and Charlie prepare their secret keys: ${K_A} = {\kern 1pt} {\kern 1pt} \{ K_A^1,{\kern 1pt} {\kern 1pt} K_A^2,{\kern 1pt} K_A^3,...,K_A^n\}$, ${K_B} = {\kern 1pt} {\kern 1pt} \{ K_B^1,{\kern 1pt} {\kern 1pt} K_B^2,{\kern 1pt} K_B^3,...,K_B^n\}$, ${K_C} = {\kern 1pt} {\kern 1pt} \{ K_C^1,{\kern 1pt} {\kern 1pt} K_C^2,{\kern 1pt} K_C^3,...,K_C^n\}$ respectively. The protocol is as follows (shown in Fig. 2).

\textbf{Step 1}. Alice (Bob, Charlie) separately prepares $n$ Bell state $\left| {{\psi ^{\rm{ + }}}} \right\rangle$ and divides the first and second qubits to form sequence ${p_A}$,(${p_B},{p_C}$) and ${q_A}({q_B},{q_C})$, respectively.

\textbf{Step 2}. Alice (Bob, Charlie) prepares $n$/2 Bell states $\left| {{\psi ^{\rm{ + }}}} \right\rangle$ as decoy qubits and concatenates them with ${q_A}({q_B},{q_C})$ to form sequence ${q'_A}$(${q'_B},{q'_C}$). Next, Alice (Bob, Charlie) applies permutation operator ${({\rm\Pi _{2n}})_A}$(${({\rm\Pi _{2n}})_B},{({\rm\Pi _{2n}})_C}$) on ${q'_A}$(${q'_B},{q'_C}$) to obtain sequence ${q''_A}$(${q''_B},{q''_C}$), the new sequence will be send to the next user Bob (Charlie, Alice).

\textbf{Step 3}. After receiving Bob's (Charlie's, Alice's) acknowledged receipt, Alice (Bob, Charlie) announces the coordinates of the decoy qubits. The receiver Bob (Charlie, Alice) performs the channel checking as same as the procedure in Step 3 of SAP1, and only if the error rate is within the tolerable limit, the protocol continues to the next step, otherwise it restarts from Step 1.

\textbf{Step 4}. After discarding decoy qubits, according to ${K_B}$(${K_C},{K_A}$), Bob (Charlie, Alice) applies unitary operation $I$ or $X$ on ${q_A}$(${q_B},{q_C}$) to obtain a new sequence ${r_B}$(${r_C},{r_A}$). Subsequently, Bob (Charlie, Alice) concatenates them with $n$/2 decoy Bell states $\left| {{\psi ^{\rm{ + }}}} \right\rangle $ and then applies the permutation operator $\left( {{\Pi _{2n}}} \right)$ to form new randomized sequence ${r''_B}$(${r''_C},{r''_A}$). Bob (Charlie, Alice) sends new sequence to the next user, Charlie (Alice, Bob).

\textbf{Step 5}. After Charlie (Alice, Bob) acknowledged the receipt of the qubits, Bob (Charlie, Alice) announces the coordinates of the decoy qubits. Each receiver performs the channel checking as same as Step 3.

\textbf{Step 6}. After discarding decoy qubits, Charlie (Alice, Bob) applies unitary operation $I$ or $Z$ on ${r_B}$(${r_C},{r_A}$) according to ${K_C}$(${K_A}, {K_B}$), and obtains a new sequence ${s_C}$(${s_A},{s_B}$). After inserting n/2 decoy Bell states $\left| {{\psi ^{\rm{ + }}}} \right\rangle $ and then applying the permutation operator $\left( {{\rm\Pi _{2n}}} \right)$, Charlie (Alice, Bob) can obtain the new sequence ${s_C}$(${s_A},{s_B}$), and sends to the next user Alice (Bob, Charlie).

\textbf{Step 7}. Charlie (Alice, Bob) and Alice (Bob, Charlie) performs the channel checking as same as Step 3.

\textbf{Step 8}. After having discarded all decoy qubits, Alice (Bob, Charlie) rearranges the received sequences and then perform Bell measurements on the particle pairs in sequences ${p_A}$(${p_B},{p_C}$) and ${s_C}$(${s_A},{s_B}$) to obtain the other two participants' secret keys. Finally, each of them can get the final key $K = {K_A} \oplus {K_B} \oplus {K_C}$.
\begin{figure}
  \centering
  \includegraphics[width=11cm]{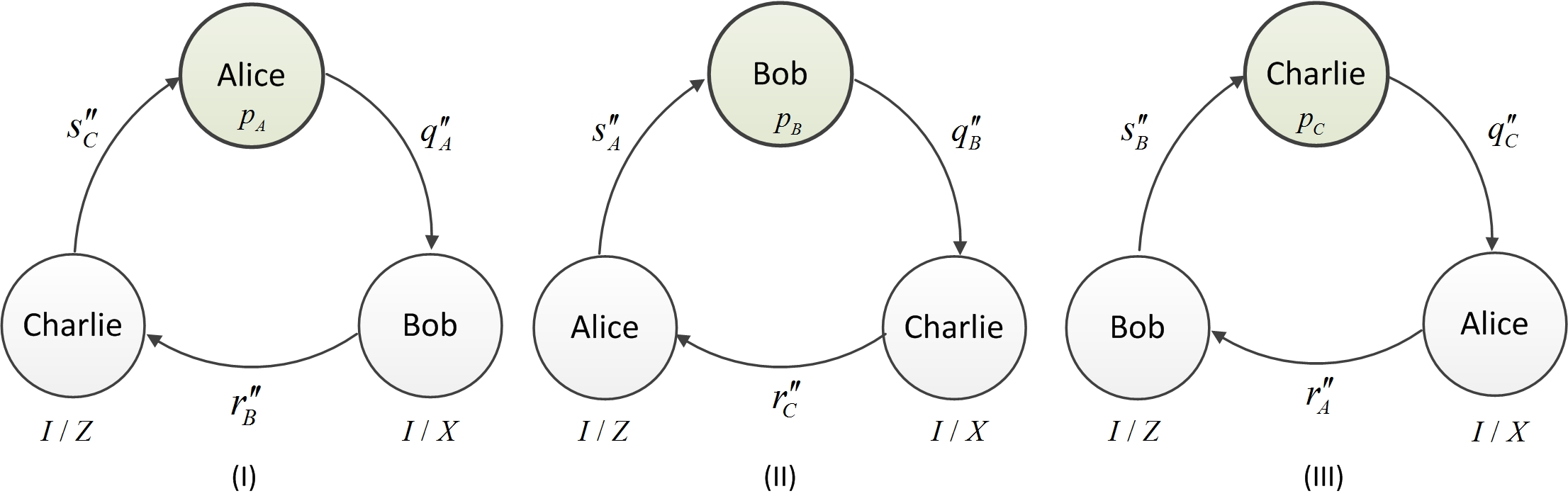}
  \caption{ The process of Shukla's three-party QKA protocol starting from Alice(I), Bob(II), Charlie(III), respectively.}
\end{figure}

\subsection{Review of Zhu et al.'s improvement}
\label{sec:2.2}

In Ref.[20], Zhu et al. pointed three security flaws in Shukla's three-party QKA protocols. First, the privacy property is not achieved in the SAP2 protocol. In Step 8 of the SAP2 protocol, the participant Alice(Bob, Charlie) performs Bell-state measurements on the corresponding particle pairs in ${p_A}$(${p_B},{p_C}$) and ${s_A}$(${s_B},{s_C}$). According to the measurement result, Alice (Bob, Charlie) can deduce the first operator applied by Bob (Charlie, Alice) and the second operator applied by Charlie (Alice, Bob). Then each user can further deduce the corresponding bits in other two participants' keys. Thus there exists privacy leakage in the SAP2 protocol. Secondly, fairness is also lost in the SAP2 protocol. i.e., two dishonest parties Alice and Bob can collaborate to determine the shared secret key. In the end of Step 5, Alice and Bob can deduce Charlie's encoding operations $\left\{ {{U_{K_C^i}}{\kern 1pt} |{\kern 1pt} {\kern 1pt} {\kern 1pt} i = 1,{\kern 1pt} 2, \cdots ,n} \right\}$ through performing Bell-state measurement on the particle pairs in ${p_B}{r_C}$.  In Step 6, Bob can choose a different unitary operation $U_i^\dag  = {U_{K_C^i}}{U_{K_B^i}}$ to perform on the $i$th particle in ${r_A}$ , which means Alice and Bob can totally offset the role of Charlie in the generation of the final key. Thus Charlie cannot equally influence the final shared key, and the SAP2 protocol cannot achieve fairness property. Finally, they pointed out there is another minor flaw in Shukla et al.'s two protocols: the eavesdropper can flip any bit of the final key without introducing any error, which can be called the eavesdropper's flip attack. Taking the SAP2 protocol as example, if an attacker performs $X$ or $Z$ on each particle in ${q''_A}$(${q''_B}, {q''_C}$) or ${r'_A}$(${r'_B},{r'_C}$), the state of each decoy photon pair does not change; however, the final states of the corresponding particle pairs in ${p_A}$(${p_B},{p_C}$) and ${s_A}$(${s_B},{s_C}$) may have been changed. In the end, Alice (Bob, Charlie) may obtain a wrong final bit through performing Bell-state measurements on the corresponding particle pairs in ${p_A}$(${p_B},{p_C}$) and ${s_A}$(${s_B},{s_C}$).

To avoid above security flaws, they proposed an improvement on the SAP2 protocol as follows. In Step 4, another additional unitary operation $I$ or $X$ is randomly chosen to perform on the $i$th particle in the received sequence. And in Step 8, Alice (Bob, Charlie) firstly announces the details of the additional unitary operation before notifying the coordinates of the message qubits, and then rearranges the sequence and then performs same additional unitary operation on the $i$th particle according to Bob's (Charlie's, Alice's) announcements. Alice (Bob, Charlie) performs Bell-state measurement on the corresponding particle pairs in ${p_A}$(${p_B},{p_C}$) and ${s_A}$(${s_B},{s_C}$) to obtain exclusive OR values of the other two participants' secret keys. In the end of these two protocols, all participants randomly choose some bits from the generated key for a final eavesdropping checking, and they announce each bit in a random sequential order. They claimed this final checking can find the eavesdropper's flip attack.

\section{Our efficient and secure $N$-party QKA protocol}
\label{sec:3}
Although Zhu et al. claimed that they solved three kinds of security flaws of Shukla et al.'s protocols, but there still exists some unsolved problems. First, the final eavesdropping checking cannot find the flip attack if the eavesdropper applies the same operations $X$ on the transmitted particles. Second, they didn't give the improvement of the two-party SAP protocol(i.e., SAP1) where there also exists the privacy leak and collusion attack. Third, similar to Shukla et al.'s protocols, each Bell state is used to carry only one classic bit, which is not efficient enough.

To solve these problems, we proposed an efficient and secure arbitrary $N$-party $\left( {N \ge 2} \right)$ QKA protocol. Suppose each user $j$  prepares its $2n$-length secret key ${K_j} = {\kern 1pt} {\kern 1pt} \{ K_j^1,{\kern 1pt} {\kern 1pt} K_j^2,{\kern 1pt} K_j^3, \cdots ,K_j^n\}$ respectively, here $1 \le j \le N$, $j + 1$ means $\left( {j + 1} \right)\bmod N$, and $K_j^i\left( {{{i = }}1,2,...,{\rm{n}}} \right)$ is randomly chosen from $\{ 00,01,10,11\}$. The whole procedures are as follows.

\textbf{Step 1}. Each user $j$ prepares $n$ Bell states $\left| {{\psi ^{\rm{ + }}}} \right\rangle$, and divides the first and the second qubits to respectively form two sequences ${p_j}$ and ${q_{j(1)}}$, here the subscript (1) represents the first participant. Then user $j$ prepares $n$ decoy photons randomly chosen from $\left\{ {\left| 0 \right\rangle ,\left| 1 \right\rangle ,\left|  +  \right\rangle ,\left|  -  \right\rangle } \right\}$,

\begin{equation}\label{1}
  \left\{ {\begin{array}{*{20}{c}}
{\left|  +  \right\rangle  = \frac{1}{{\sqrt 2 }}(\left| 0 \right\rangle  + \left| 1 \right\rangle )}\\
{\left|  -  \right\rangle  = \frac{1}{{\sqrt 2 }}(\left| 0 \right\rangle  - \left| 1 \right\rangle )}
\end{array}} \right.
\end{equation}
and randomly insert them into ${q_{j(1)}}$ to form a new sequence ${q'_{j(1)}}$. Then each user $j$ sends sequence ${q'_{j(1)}}$ to the next user $j+1$.

\textbf{Step 2}. After receiving the acknowledged receipt from user $j+1$, user $j$ announces the positions and the measurement basis of the decoy qubits, and then user $j+1$ performs the channel checking by measuring the decoy qubits. If the error rate is within tolerable limit, the protocol continues the next step; otherwise it goes back to Step 1.

\textbf{Step 3}. After discarding the decoy qubits, user $j+1$ applies the unitary operations $\left\{ {I,{{ Z}},{{ X}},{{ Y}}} \right\}$ on the remaining ${q_{j(1)}}$ sequence in accordance with his key ${K_{j + 1}}$. That is, if $K_{j + 1}^i$=00 (01, 10, 11), he applies $I({{Z}},{{ X}},{{ Y)}}$ on the corresponding target qubit (also shown in Table 1).

\begin{equation}\label{1}
  \left\{ {\begin{array}{*{20}{c}}
\begin{array}{l}
I = \left| 0 \right\rangle \left\langle 0 \right| + \left| 1 \right\rangle \left\langle 1 \right| = \left[ {\begin{array}{*{20}{c}}
1&0\\
0&1
\end{array}} \right]\\
Z = \left| 0 \right\rangle \left\langle 0 \right| - \left| 1 \right\rangle \left\langle 1 \right| = \left[ {\begin{array}{*{20}{c}}
1&0\\
0&{ - 1}
\end{array}} \right]
\end{array}\\
X = \left| 0 \right\rangle \left\langle 1 \right| + \left| 1 \right\rangle \left\langle 0 \right| = \left[ {\begin{array}{*{20}{c}}
0&1\\
1&0
\end{array}} \right]\\
{Y = \left| 0 \right\rangle \left\langle 1 \right| - \left| 1 \right\rangle \left\langle 0 \right| = \left[ {\begin{array}{*{20}{c}}
0&1\\
{ - 1}&0
\end{array}} \right]}
\end{array}} \right.
\end{equation}
As a result of encoding operations, user $j+1$  obtains a new sequence ${q_{j(2)}}$. Then user  $j+1$ randomly inserts $n$ decoy single photons into ${q_{j(2)}}$ to form a new randomized sequence ${q'_{j(2)}}$, and then sends it to the next user $j+2$.

\begin{table}[htbp]
  \centering
  \caption{The unitary operation according to the secret key}
    \begin{tabular}{cc}
    \hline
    Secret key & Unitary operation  \bigstrut\\
    \hline
    00    & \textit{I} \bigstrut[t]\\
    01    & \textit{Z} \\
    10    & \textit{X} \\
    11    & \textit{Y} \bigstrut[b]\\
    \hline
    \end{tabular}%
  \label{tab:addlabel}%
\end{table}

\textbf{Step 4}. After receiving the acknowledged receipt from user $j+1$, the receiver $j+2$ performs the channel checking as same as Step 2.

\textbf{Step 5}. If the receiver is exactly the initial sender (i.e., $(j+2)$mod $N$ = $j$, user $j+2$ is exactly user $j$), the protocol continues Step 6. Otherwise, it goes to Step 3 and Step 4, i.e., user $j+2$ performs encoding procedure in Step 3, and user $j+3$ performs the channel checking in Step 4.

\textbf{Step 6}. After all users finished the above encoding and transmission procedures, the sequences have been transmitted back to the initial senders. Now each user $j$ has two ordered sequences $p_j^{}$  and ${q_{j(N - 1)}}$. Each user $j$ performs Bell measurements on $p_j^{}{q_{j(N - 1)}}$. According to the measurement result, each user can obtain the secret keys of the other $N$-1 parties. Hence, they shares the secret key $K = {K_1} \oplus {K_2} \oplus  \cdots  \oplus {K_N}$.

\begin{figure}
  \centering
  \includegraphics[width=9cm]{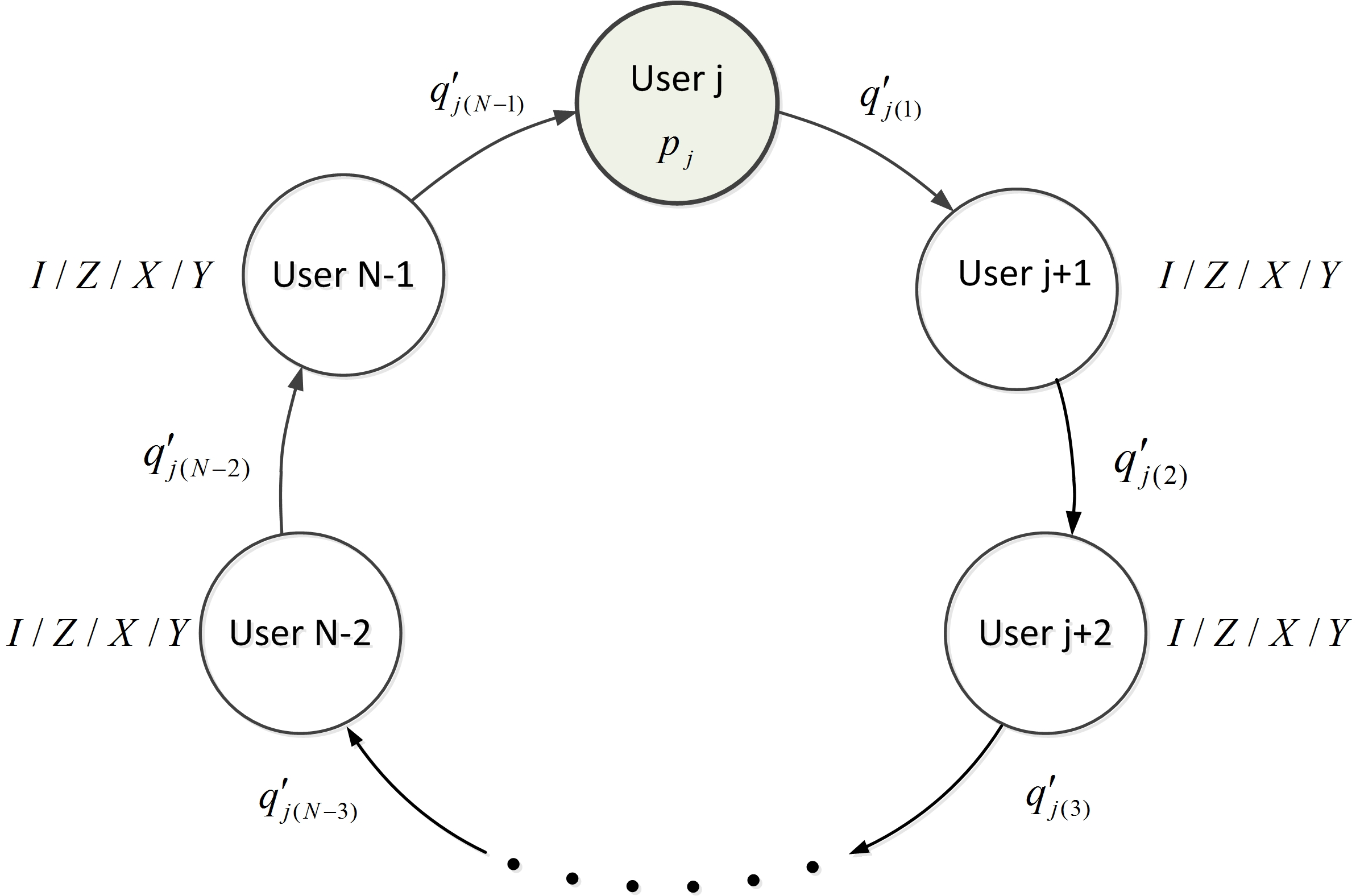}
  \caption{Schematic diagram of our $N$-party Quantum key agreement protocol}
\end{figure}

It should be noticed that the ``post-measurement'' mechanism is introduced to prevent the collusion attack in the above protocol. To be specific, the measurements of Step 6 will be performed only after all the users have finished his own transmission round and all the message qubits have returned back to the corresponding initial senders.

In addition, since our proposed protocol is an arbitrary $N$-party QKA protocol, it includes the two-party ($N$=2) and three-party ($N$=3) cases. The two-party protocol(shown in Fig. 4) can be considered as the improvement of the SAP1 protocol, and the three-party protocol(shown in Fig. 5) is the improvement of the SAP2 protocol. And Table 2 shows the transformation of message state $\left| {{\psi ^ + }} \right\rangle$ in the three-party protocol.

\begin{figure}
  \centering
  \includegraphics[width=11cm]{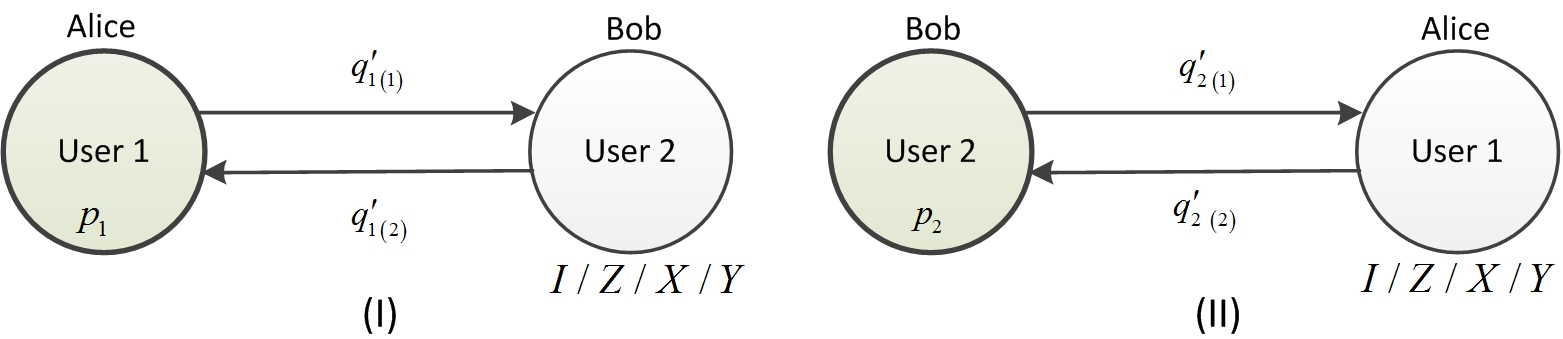}
  \caption{The process of two-party scenario for our QKA protocol. (I) Alice sends sequence ${q'_{1(1)}}$ to Bob, then Bob encodes his key by applying ${I/Z/X/Y}$ on each qubit in sequence ${q_{1(1)}}$, and Bob sends back sequence ${q'_{1(2)}}$ to Alice. (II) At the same time, Bob also sends sequence ${q'_{2(1)}}$ to Alice for encoding operations, and then receives sequence ${q'_{2(2)}}$ from Alice. After Alice and Bob both received sequences $q'_{1(2)}$, $q'_{2(2)}$ from each other, they separately perform the Bell measurement and obtain each other's key.}
\end{figure}

\begin{figure}
  \centering
  \includegraphics[width=11cm]{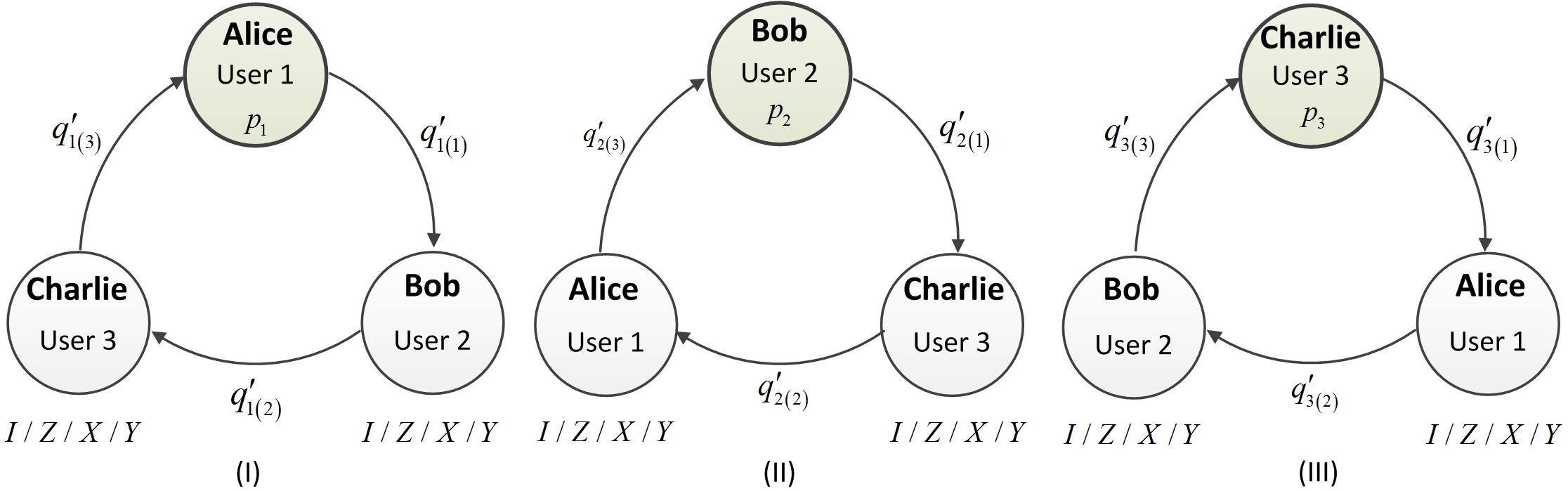}
  \caption{The process of three-party scenario for our QKA protocol. Here (I),(II),(III) represent the corresponding round beginning from the sender Alice, Bob and Charlie, respectively. The detailed procedures of each round are similar to the two-party scenario.}
\end{figure}

\begin{table}[htbp]
  \centering
  \caption{Transformation of $\left| {{\psi ^ + }} \right\rangle $ in the three-party scenario of our protocol}
    \begin{tabular}{l|llll}
    \hline
    \multicolumn{1}{l|}{\multirow{3}[2]{1.6cm}{Initial state prepared by user $j$}} & \multicolumn{1}{l}{\multirow{3}[2]{2.2cm}{key of user $j+1$ and its operations}} & \multicolumn{1}{l}{\multirow{3}[2]{2.2cm}{key of user $j+2$ and its operations}} & \multicolumn{1}{l}{\multirow{3}[2]{2.2cm}{Exclusive OR between user $j+1$ and $j+2$}} & \multicolumn{1}{l}{\multirow{3}[2]{*}{Final state}} \bigstrut[t]\\
    \multicolumn{1}{c|}{} &       &       &       &  \\
    \multicolumn{1}{c|}{} &       &       &       &  \bigstrut[b]\\
    \hline
    \multirow{16}[8]{*}{$\frac{{\left| {00} \right\rangle  + \left| {11} \right\rangle }}{{\sqrt 2 }}$} & 00($I$) & 00 ($I$) &       & \multirow{4}[2]{*}{} \bigstrut[t]\\
          & 01($Z$) & 01($Z$) & \multicolumn{1}{l}{00} & $\frac{{\left| {00} \right\rangle  + \left| {11} \right\rangle }}{{\sqrt 2 }}$ \\
          & 10($X$) & 10($X$) &       &  \\
          & 11($Y$) & 11($Y$) &       &  \bigstrut[b]\\
\cline{2-5}          & 00($I$) & 01($Z$) &       & \multirow{4}[2]{*}{} \bigstrut[t]\\
          & 01($Z$) & 00($I$) & \multicolumn{1}{l}{01} & $\frac{{\left| {00} \right\rangle  - \left| {11} \right\rangle }}{{\sqrt 2 }}$ \\
          & 10($X$) & 11($Y$) &       &  \\
          & 11($Y$) & 10($X$) &       &  \bigstrut[b]\\
\cline{2-5}          & 00($I$) & 10(X) &       & \multirow{4}[2]{*}{} \bigstrut[t]\\
          & 01($Z$) & 11($Y$) &       &  \\
          & 10($X$) & 00($I$) & \multicolumn{1}{l}{10} & $\frac{{\left| {01} \right\rangle  + \left| {10} \right\rangle }}{{\sqrt 2 }}$ \\
          & 11($Y$) & 01($Z$) &       &  \bigstrut[b]\\
\cline{2-5}          & 00($I$) & 11($Y$) &       & \multirow{4}[2]{*}{} \bigstrut[t]\\
          & 01($Z$) & 10($X$) &       &  \\
          & 10($X$) & 01($Z$) & \multicolumn{1}{l}{11} & $\frac{{\left| {01} \right\rangle  - \left| {10} \right\rangle }}{{\sqrt 2 }}$\\
          & 11($Y$) & 00($I$) &       &  \bigstrut[b]\\
    \hline
    \end{tabular}%
  \label{tab:addlabel}%
\end{table}%

\section{Security and efficiency analysis}
\label{sec:4}
\subsection{Security analysis}
\label{sec:4.1}
In Ref.[28], Shukla et al. have already discussed the unconditional security of the eavesdropping checking with Bell states to resist the external attacks. In our protocol, we introduced the eavesdropping checking with decoy single photons instead of Bell states. In essence, the eavesdropping checking strategy plays the same role against the external attacks as that in Shukla et al.'s protocols. So, We omit the process of analyzing the security property of the protocol, and mainly focus on the correctness, privacy and fairness aspects.

We firstly consider the correctness property. Without loss of generality, we suppose the outside eavesdropper Eve launches the flip attack, and tries to flip the bits of the final key without being detected. Specifically, Eve performs $X$ or $Z$ on each particle in ${q'_{j(1)}}$ (${q'_{j(2)}}$, ..., ${q'_{j(N-1)}}$), $1\leq j \leq N-1$, which will change the final states of the corresponding particle pairs in ${p_1}$(${p_2}, ..., {p_N}$) and ${q_{1(N-1)}}$(${q_{2(N-1)}}$, ..., ${q_{N(N-1)}}$). However, the action will soon be found in the next process of eavesdropping checking because we replace the original Bell states with single photons (the $X$ or $Z$ operation will change the states of decoy single photons). So, the single photon checking strategy can avoid the flip attack, and thus guarantees the correctness property of the protocol.

Secondly, the privacy property of our protocol can be achieved by applying four pauli operations in the encoding phase. For convenience, we take the three-party scenario of our protocol as example, and suppose a vicious party want to deduce the others' secret keys after the Bell-state measurements on the final states they received. In Table 2, we have summarized all the possible cases of the message state $\left| {{\psi ^ + }} \right\rangle$ in the whole protocol. As shown in the table, the vicious party can only get the exclusive OR value of all the other participants' keys (the fourth column of Table 2) from the Bell-state measurement result (the fifth column of Table 2), and cannot deduce any information about the specific keys of other participants. For instance, we suppose vicious Alice get the measurement result $\left| {{\psi ^ - }} \right\rangle  = \frac{{\left| {00} \right\rangle  - \left| {11} \right\rangle }}{{\sqrt 2 }}$, and try to get Bob's or Charlie's key. Through comparing the final state $\left| {{\psi ^ - }} \right\rangle$ with the initial state $\left| {{\psi ^ +}} \right\rangle$, Alice can deduce that the exclusive OR value of Bob's and Charlie's keys is 01. However, the secret value of Bob(Charlie) may be 00(01), 01(00), 10(11), or 11(10), i.e., there exists all the combination possibilities with equal probability. So, it is obviously impossible for Alice to exactly deduce Bob's or Charlie's secret key.

The inner participants' collusion attack is one of the most threatening attacks against the fairness property of QKA. Specifically, some malicious participants may cooperate with each other, and perform some operations to offset the other participants' keys(encoding operations) as mentioned in Ref.[20]. Also Taking the three-party protocol as example, we suppose Charlie is honest and the dishonest parties Alice and Bob want to determine the shared key alone. The most possible way is that Alice and Bob cooperate with each other and try to deduce $K_C$ in the first round starting from Alice (see Fig. 5(I)). In this round, Alice and Bob firstly cooperate to get $K_B$, $K_A$ from each other, and then Alice performs Bell-sate measurement on $p_j$$q'_{1(3)}$ after receiving sequence $q'_{1(3)}$ from Charlie. From the measurement results, Alice can get the exclusive OR value between $K_B$ and $K_C$, i.e., $K_{BC} = K_{B} \bigoplus K_{C}$. Since Alice has known the value of $K_B$, so it is easy for her to get the secret key of Charlie without being detected: $K_C$=$K_{BC}\bigoplus K_{B}$. In order to offset the affection from Charlie and determine the shared key alone, Alice and Charlie will change the encoding operations in the following second round (Fig. 5(II)) and the third round (Fig. 5(III)). For example, Alice may perform the operations according to $K_A \bigoplus K_{C}$ instead of $K_A$ in the second round beginning from Bob, and Bob performs the corresponding operations according to $K_B \bigoplus K_{C}$ instead of $K_B$ in the third round beginning from Charlie. In this way, maybe the shared key can be just determined by the dishonest Alice and Bob, i.e., $K$=$K_C \bigoplus (K_A \bigoplus K_{C})\bigoplus K_B = K_A \bigoplus K_B$ for the second round, $K$=$K_A \bigoplus (K_B \bigoplus K_{C})\bigoplus K_C = K_A \bigoplus K_B$ for the third round. However, In our protocol, the post-measurement mechanism is introduced to prevent against this kind of security flaws. As described in Step 6 of our protocol, each participant will perform the Bell-state measurements only after all the other participants have also finished their own rounds and the message qubits have returned back to them. In other words, all the three rounds are running simultaneously in the three-party protocol, and Alice, Bob and Charlie will perform the Bell measurement at the same time. In this mechanism, the dishonest Alice and Bob have no chance to change the operations in the second and third round according to Alice's measurement results in the first round. So, we can say our protocol can prevent against this kind of collusion attack, which guarantees the fairness property of QKA.

\subsection{Effiency analysis}
\label{sec:4.2}

According to Ref.[29], the well-known efficiency measurement of secure quantum communication is as follows:

\begin{equation}\label{3}
  \eta  = \frac{c}{{q + b}}
\end{equation}
where $c$ denotes the total number of transmitted classical bits (message bits), $q$ represents the total number of qubits used in the protocol, and $b$ denotes the whole classical bits exchanged in the protocol.

Taking into account that other analogous QKA protocols are in the form of three parties, we take the three-party ($N$=3) scenario of our protocol to evaluate the efficiency. In the protocol, four Pauli operations $\left\{ {I,{\rm{ }}Z,{\rm{ }}X,{\rm{ }}Y} \right\}$ are used to encode two bits in the encoding phase, that means the message bits is $c$=$2n$. And each party prepares $n$ Bell states as quantum resource and $3n$ single photons are utilized for eavesdropping checking, thus $q$=$3\times(2n+3n)$ = $15n$. In addition, each party uses $3n$ bits of classical information to disclose the coordinates of the message qubits, so the exchanged classic bits $b$=${3 \times 3n = 9n}$. Thus, the efficiency of our three-party QKA protocol is ${\eta _{our}} = \frac{{2n}}{{15n + 9n}} = 8.33\%$.

In Shukla's three-party QKA protocol(SAP2), the values of $q$ and $b$ are also $15n$ and $9n$, respectively. The difference is that their message bits are only $n$ bits, thus the efficiency is ${\eta_{shukla}}=\frac{n}{{15n + 9n}}=4.17\%$. In Zhu et al.'s improved protocol, the values of $c$ and $q$ are the same as Shukla's three-party protocol, i.e., $c$=$n$, $q$=$15n$. However, Since each user needs to announce the classic message of his additional unitary operations to the initial sender, so the additional exchanged classic bits are $N$ bits in a round, and the total exchanged classic bits is $b = 3 \times (3n + n)= 12n$. Thus, the efficiency is ${\eta _{zhu}}$ = $\frac{n}{{15n + 12n}}$ = $3.57\% $. For the sake of convenience, Table 2 lists the relevant comparison results among the three protocols.
\begin{table}[htbp]
  \centering
  \caption{Comparison among our protocol and the other analogous three-party protocols}
    \begin{tabular}{lllll}
    \hline
    protocols & message bits & qubits & exchanged bits & efficiency coefficient \bigstrut\\
    \hline
    SAP2  & $n$     & $15n$   & $9n$    & 4.17\% \bigstrut[t]\\
    Zhu et al.'s  & $n$     & $15n$   & $12n$   & 3.57\% \\
    Our   & $2n$    & $15n$   & $9n$    & 8.33\% \bigstrut[b]\\
    \hline
    \end{tabular}%
  \label{tab:addlabel}%
\end{table}%

Besides the higher efficiency coefficient, the other advantage is that we employ the single photon for channel checking instead of Bell states. As we know, the single photon is easier to be prepared than Bell states in practice, so our protocol seems more feasible and economic.

\section{ conclusion}
In this study, we proposed an arbitrary $N$-party quantum key agreement protocol, which includes the improvement of the Shukla's two QKA protocols, also shows better performance than the Zhu et al's improvement. In our protocol, four Pauli operations $\left\{ {I,{\rm{ }}Z,{\rm{ }}X,{\rm{ }}Y} \right\}$ are used in the encoding phase instead of the original two operations $\left\{ {I,{\rm{ }}X} \right\}$ , which not only enhances the protocol's efficiency, but also protects the participant's secret key being deduced by the other participants and then guarantees the privacy property of QKA. We introduce the single photon checking strategy to avoid eavesdropper's flip attack, and thus achieves the correctness property. What is more, the post-measurement mechanism guarantees that no participant in our protocol can offset the other's operations through performing the collusion attack, so the fairness property is achieved too.

With the popularity of cloud platforms, more and more sensitive or private information has been outsourced onto the cloud, so cloud secure storage[30,31] and searchable encryption on cloud data[32-34] become a big concern.
Compared with QKD, QKA is more suitable for such distributed systems. It would be interesting to explore the possibility of designing quantum counterparts based on QKA for the classical schemes designed in the context of encrypted cloud data[35-38].

\begin{acknowledgements}
The authors would like to thank the anonymous reviewers and editor for their comments that improved the quality of this paper. This work is supported by the National Nature Science Foundation of China (Grant Nos. 61502101, 61501247 and 61672290), the Six Talent Peaks Project of Jiangsu Province (Grant No. 2015-XXRJ-013), Natural Science Foundation  of Jiangsu Province(Grant Nos. BK20171458, BK20140823), Natural science Foundation for colleges and universities of Jiangsu Province(Grant No.16KJB520030), and the Priority Academic Program Development of Jiangsu Higher Education Institutions(PAPD).
\end{acknowledgements}



\end{document}